\documentstyle[12pt]{article}
\newcommand{\be}{\begin{equation}}
\newcommand{\ee}{\end{equation}}
\newcommand{\ed}{\end{document}}
\newcommand{\lab}[1]{\label{#1}}
\newcommand{\re}[1]{(\ref{#1})}
\newcommand{\ci}[1]{\cite{#1}}
\renewcommand{\baselinestretch}{1.4}
\title{DIRAC ELECTRON IN THE ELECTRIC DIPOLE FIELD}
\author{V.I.MATVEEV, M.M.MUSAKHANOV\\
Department of Theoretical Physics, Tashkent State University \\
700095, Vuzgorodok, Tashkent, Uzbekistan \\
e-mail: yousuf@aphi.silk.glas.apc.org \\
and \\
D.U.MATRASULOV \\
Thermal Physics Department of Uzbek Academy of Sciences\\
700135, ul.Katartal 28, Tashkent, Uzbekistan\\ }
\begin{document}
\begin{titlepage}
\maketitle
\begin{abstract}

  Dirac equation for the finite dipole potential is solved by
the method of the join of the asymptotics. The formulas for the near
continuum state energy term of a relativistic electron-dipole system
are obtained analytically. Two cases are considered: $Z < 137$
and $Z >137$

  Dirac equation for $Z > 137$ is solved by usual method of cut-off
potential at small distances.
\end{abstract}
\end{titlepage}

   We would like to discuss the problem of the Dirac equation
for an electron in an electric dipole field which is closely related
to the problem of exotic heavy quarkonium states too. We mean the
problem of the motion of light valence and vacuum quarks in the
color field of heavy quark-antiquark pair.
  The motion of an electron under the influence of an electric
dipole one of the classical problem of nonrelativistic quantum
mechanics. Energy eigenvalues of such system was found by many authors
see e.g. \ci{1,2,3}. Several authors have calculated the critical
value of the dipole moment at which energy term reaches the
boundary of continuum \ci{1,2}.

  Solution of corresponding relativistic problem is more difficult
mathematical problem, since variables of the Dirac equation can't be
separated at any orthogonal system of coordinates. Near continuum
energy state  of the relativistic electron, moving in the field
of finite electric dipole, we shall find by the method of the join
of the asymptotics which successfully was applied to solution of
two-center Coulomb problem for the Dirac equation (with same signs
of charges)\ci{6,7}.
Previously this problem was considered in our paper \ci{8}.

  In the present paper we also use asymptotical separability of
variables at large distances in the relativistic two-center Coulomb
problem and additionally discuss the nonrelativistic limit of our results.
Then the squared Dirac equation takes a form of the
Schrodinger equation at which variables can be separated.
Let us consider relativistic electron moving in the field of a
finite electric dipole which is composed of charges
$+Z\alpha$ and $-Z\alpha$
are separated by distance R. The potential of this system is given by
 \be V = \frac{Z\alpha}{r_{1}}-\frac{Z\alpha}{r_{2}}\ee

where $r_{i}$  is the distance from $i$ th charge,
$\alpha^{-1} = 137$   (the system of
units $\hbar = m = c = 1$ is used here and below).
The motion of this  electron is described by Dirac equation which
can be written as \be i \frac{\partial\psi}{\partial t} = H\psi\;,\ee

where $H=\vec{\alpha}\vec{p}+\beta+V$ is the Dirac Hamiltonian,
$\alpha$ and $\beta$ are the Dirac matrices.

  In the spinor representation stationar squared equation can be
written in the components $\psi_{1}$ and $\psi_{2}$ of a
byspinor $\psi=\left(\begin{array}{c} \psi_{1} \\
\psi_{2} \end{array} \right)$ as in [5]
\be [(\varepsilon-V)^{2}+ \triangle +
i\vec{\sigma}\vec{\nabla}V-1]\psi_{1}\ee
\be [(\varepsilon-V)^{2}+ \triangle +
i\vec{\sigma}\vec{\nabla}V-1]\psi_{2} \ee
where $ \varepsilon$ is energy of an electron, $\vec{\sigma}$
are standard Pauli matrices.

   At large distances from dipole we can neglect by terms $V^{2}$  and
$i\vec{\sigma}\vec{\bigtriangledown}V$ with respect to V hence for each
component we have equation
\be [\triangle - 2\varepsilon V + \varepsilon^{2}-1]\psi = 0
\lab{e1}
\ee

  Eq.\re{e1} is the Schrodinger equation with potential
$\varepsilon V$ and
with energy $ -\frac{1}{2}(\varepsilon^{2} -1)$
  So, at large distances the wave function has separated form
 \be \psi=\frac{U(\xi)}{(\xi^{2}-1)^{1/2}}\,
\frac{V(\eta)}{(1-\eta^{2})^{1/2}}\,
 exp(im\varphi) \ee
 where $\xi$, $\eta,$ $\varphi$ are the spheroidal coordinates
which are defined by
 $\xi=(r_{1}+r_{2})R^{-1},$ $\eta=(r_{1}-r_{2})R^{-1},$
$\varphi=arctg(\frac{y}{x}),$
 where $y$ and $x$ are the cartesian coordinates,
R is the distance between charges,
 $ 1<\xi<\infty\,,\;\;-1\leq\eta\leq1\,,\;\;0\leq\varphi\leq2\pi$

The radial and angular equations become
\be U^{\prime\prime}(\xi)\; +\; [-p^{2}\; +\; \frac{A}
{1-\xi^{2}}\, +\,\frac{1-m^{2}}
{(1-\xi^2)^{2}}]U(\xi)\,=\, 0
\lab{rad}
\ee
\be V''(\eta)\;+\; [-p^{2}\;+\; \frac{D\eta\, -\,A}
{1\,-\,\eta^{2}}\; +\; \frac{1\,-\,m^{2}}
{(1\,-\,\eta^{2})^{2}}]V(\eta)\;=\; 0
\lab{ang}
\ee
where $p^{2}=-\frac{R^{2}}{4}(\varepsilon^{2}-1),\;
D=2\mid\varepsilon\mid RZ\alpha,\:$
 A is the constant of separation.

The boundary conditions are
\be U(1)=0,\;U(\xi)_{\xi\longrightarrow\infty}\longrightarrow 0,\;
V(\pm 1)=\;0 \ee

Asymptotes of eq.\re{rad} for large $\xi$ is [1] $U(\xi)\;=
\;(p\xi)^{1/2} K_{i\nu}(p\xi),$
where $K_{i\nu}$ is the McDonald function, $\nu^{2}=A - \frac{1}{4}.$
Solution of eq.\re{ang} in the vicinity of $ \eta = 0 $
was also found in ref.[1].

This asymptotes has the form
\be V(\eta)\;=\; \left[z'(\eta)\right]^{-\frac{1}{2}}
Ai(-D^{\frac{1}{3}}z)\ee
where Ai is the Airey function,
$z(\eta)=[\frac{3}{2}\int^{\eta}_{0}\eta^{1/2}
(1-\eta^{2})^{-1/2}d\eta]^{\frac{2}{3}}$ and
$z'(\eta)=\frac{dz}{d\eta}.$

Thus for large $p\xi $ and for small $\eta $
asymptotes of wave function
can be written in the form(for $m=0 $)
\be \psi = [\frac{\pi}{\nu sh\pi\nu}]^{\frac{1}{2}}(p\xi)^
{\frac{1}{2}}(\xi^{2}-
1)^{-\frac{1}{2}}Sin(\nu ln[\frac{2}{p\xi} + arg\gamma(1+i\nu)])
\lab{sm}
\ee

Here we have used asymptotes of function $K_{i\nu}(x)$ for
$x\longrightarrow 0 $
\ci{6}
$$K_{i\nu}(x) \longrightarrow [\frac{\pi}{\nu sh\pi\nu}]^
{\frac{1}{2}}(p\xi)^{\frac{1}{2}}
 sin\nu ln[\frac{2}{x} + arg\gamma(1+i\nu)]$$ and
the asymptotes of Airey function
 \ci{15} $Ai(z)\longrightarrow c_{1}- c_{2}z$ for
$z\longrightarrow 0,$ where
 $c_{1}=0.355, c_{2}= 0.259,$ and the relation $z'(\eta)\approx 1$
for $\eta\longrightarrow 0.$

Following by Popov \ci{6} as an asymptotes of wave function
at small distances
we will take the product of relativistic one-center wave functions.
Here we must consider two cases:  $Z<137$ and $Z >137.$

Let $Z<137.$ In this case asymptotics (at small
distances from center) is given by [5]
\be \phi_{1} = r^{-1+\gamma} \ee
where $\gamma =(1-Z^{2}\alpha^{2})^{\frac{1}{2}}$

Thus asymptotes of two center wave function at small distances
from dipole can be written as follows:
\be \psi_{1} \sim (r_{1}r_{2})^{\gamma-1}=
(\xi^{2}-\eta^{2})^{\gamma-1} \ee

Let now $Z >137.$ In this case energy becomes imaginary
("fall to the center")
\ci{9,10}. In order to solve the Dirac equation
in this case we must cut-off the
Coulomb potential at small distances \ci{11,12}.

General form of the cut-off potential is \ci{13,14}
$$
V(r)=\left\{\begin{array}{ll}\frac{Z\alpha}{r}\,, & for\;\:r>b \\
\frac{Z\alpha}{b}f(r)\,, & for\:\;0<r<b\end{array} \right.\,, $$
where b is the radius of the nuclei.

For $f(r)=1$ (surface distribution of charge) asymptotes
of wave function has the form\ci{13}
\be \phi_{2} \sim r \ee
Hence for the case of $Z >137$ asymptotics
of two center wave function at
small distances from dipole can be written in the form
\be \psi \sim r_{1}r_{2}\sim \xi^{2}-\eta^{2} .\ee
So in general case one can write for the asymptotics of wave function
at small distances from dipole (for $R<< 1$)
\be \psi_{1} \sim (\xi^{2}- \eta^{2})^{\beta},\ee
where
$$
\beta=\left\{\begin{array}{ll}\gamma-1\,, & for\;\:Z>137 \\
1\,, & for\:\;Z>137\end{array} \right.\,, $$

For small $\eta$
\be \psi_{1} \sim \xi^{2\beta}
\lab{larg}
\ee
Since $\mid \varepsilon\mid \approx 1$ asymptotes of wave function at
small and at large distances join with other in the region $1<<\xi <<
\frac{1}{p}.$ Equating logarithmic derivatives of functions \re{larg}
and \re{sm}
and taking into account that in this region $ln\frac{2}{p\xi}\approx
ln\frac{2}{p}$ we have
\be \frac{1}{2} + \nu ctg[\nu ln\frac{2}{p}+arg\Gamma(1+i\nu)]=
-2\beta \ee
 From this expression we get
\be p = exp(\frac{1}{\nu}arg\Gamma(1+i\nu)-\frac{1}{\nu}
arcctg[-\frac{1+4\beta}
{2\nu}]-2ln2)
\lab{exp}
\ee

In order to find the $\varepsilon(R)$ from \re{exp} we must
know the parametr $\nu.$
Here we use results of ref.\ci{1} where dipole moment was found as a
function of $\nu.$  In application to our problem for ground state this
formula can be written as
\be D= \frac{\Gamma^{4}(\frac{1}{4})}{32\pi}[1+\frac{92}
{3\pi}(4p^{2}+6\nu^{2}
-1)] \ee
where $D=2\mid\varepsilon\mid ZR\alpha.$ From this equation
we find $\nu:$
\be \nu=[\frac{8\pi^{2}}{\Gamma^{4}(\frac{1}{4})}[D-2\mid\varepsilon
\mid ZR\alpha-D_{cr}]-\frac{R^{2}}{6}(\varepsilon^{2}-1)]^{\frac{1}{2}}
\lab{ptt}
\ee
where
$D_{cr}= \frac{\Gamma^{4}(\frac{1}{4})}{32\pi}[1-\frac{2}{3\pi}].$

is the critical value of dipole moment at which term
$\varepsilon(R)$ reaches
the boundary of continuum.
Taking into account relation between p and $\varepsilon$ we find
\be \varepsilon_{\pm}=\pm[1-\frac{2}{R^{2}}exp(\omega(\nu,Z))] \ee
where
\be \omega(\nu,Z)=\frac{2}{\nu}arg\Gamma(1+i\nu)-\frac{2}{\nu}arcctg
[-\frac{1+4\beta}{2\nu}]-2ln2 \ee
 Thus for $Z < 137$  we have transcendental equation for $\varepsilon(R)$
 \be \varepsilon_{\pm}=\pm 1\mp \frac{2}{R^{2}}
exp[\frac{2}{\nu}arg\Gamma(1+i\nu)-
 \frac{2}{\nu}arcctg\frac{3-4\gamma}{2\nu}-2ln2]
 \lab{smz}
 \ee
where $\nu$ defined by (21).

For $Z < 137\sqrt{7}/4 $ eq.\re{smz} can be solved by
iterations (since iteration
procedure converges only for $3 - 4\gamma < 0$ )
So for the first iteration we have
\be
\varepsilon_{\pm}=\pm 1\mp
\frac{2}{R^{2}}
exp[-\frac{\Gamma^{2}(\frac{1}{4})}{\pi(4Z\alpha
R-2D_{cr})^{\frac{1}{2}}}
arcctg\frac{\Gamma^{2}(\frac{1}{4})(3-4\gamma)}
{4\pi(4Z\alpha R-2D_{cr})^{\frac{1}{2}}}
\lab{term1}
\ee

  Formula \re{term1} is the near continuum state energy
term of an electron in the
electric dipole field. As well known nonrelativistic
limit gets near the
continuum. Since for near continuum states
$4ZaR - 2D_{cr}\sim 0 ,\; arcctg
\longrightarrow \pi .$  Thus in order to obtain
nonrelativistic limit we must
do following replacements: 1)$\gamma\longrightarrow 1,$
2) $arcctg \longrightarrow\pi$; as a result we have
\be
\varepsilon_{\pm}=\pm 1\mp \frac{2}{R^{2}}
exp[-\frac{\Gamma^{2}(\frac{1}{4}}
{\pi(4Z\alpha R-2D_{cr})^{\frac{1}{2}}}]
\lab{nonr}
\ee
 This formula for $\varepsilon_{\pm}$ coincides with
known formula from \ci{2}
 for the nonrelativistic electron-dipole system.
 By expanding arccotangent into series in formula
\re{term1} we calculate the
 correction to eq.\re{nonr}
\be
\varepsilon\approx \frac{2}{R^{2}}exp[-\frac{\Gamma^{2}(\frac{1}{4}}
{\pi(4Z\alpha R-2D_{cr})^{\frac{1}{2}}}-\frac{4}{3-4\gamma}]\ee
 From this formula it is easy to see that nonrelativistic limit take
place for           $(4Z\alpha R -2D_{cr})^{\frac{1}{2}} << 0.8 .$
 We note that this nonrelativistic limit can be also obtained directly
 from eq.\re{exp}.
  For $Z > 137$  we have equation
\be \varepsilon_{\pm}=\pm 1\mp \frac{2}{R^{2}}exp[\frac{2}{\nu}
arg\Gamma(1+i\nu)-
\frac{2}{\nu}arcctg\frac{5}{2\nu}-2ln2-\frac{2\pi}{\nu}] \ee
Solving this equation by iterations for the first iteration we have
\be
\varepsilon_{\pm}\approx \pm 1\mp \frac{2}{R^{2}}
exp[-\frac{\Gamma^{2}(\frac{1}{4})}
{\pi(4Z\alpha R-2D_{cr})^{\frac{1}{2}}}
(arcctg\frac{5\Gamma^{2}(\frac{1}{4})}
{4\pi(4Z\alpha R-2D_{cr})^{\frac{1}{2}}-\pi})]
\ee
Derived analytical formulas have to be useful for further numerical
calculations in nonasymptotical region.

We are planning to generalize our results for the Coulomb plus
confinement potential case for the calculations of the properties
of the exotic quarkonium states.

\section*{Acknowledgements}

This work was supported in part by the Foundation for
Fundamental Investigations
of Uzbekistan State Committee for Science and Technics under
contract $N$ 40,
by International Science(Soros) and INTAS foundations grants.

\ed